\input jytex.tex   
\typesize=10pt \magnification=1200 \baselineskip17truept
\footnotenumstyle{arabic} \hsize=6truein\vsize=8.5truein

\sectionnumstyle{blank}
\chapternumstyle{blank}
\chapternum=1
\sectionnum=1
\pagenum=0

\def\begintitle{\pagenumstyle{blank}\parindent=0pt
\begin{narrow}[0.4in]}
\def\endtitle{\end{narrow}\newpage\pagenumstyle{arabic}}


\def\beginexercise{\vskip 20truept\parindent=0pt\begin{narrow}[10
truept]}
\def\endexercise{\vskip 10truept\end{narrow}}


\def\eql#1{\eqno\eqnlabel{#1}}
\def\ref{\reference}
\def\peq{\puteqn}
\def\pref{\putref}

\def\mgn{\marginnote}
\def\bex{\begin{exercise}}
\def\eex{\end{exercise}}


\font\open=msbm10 


\def\StretchRtArr#1{{\count255=0\loop\relbar\joinrel\advance\count255 by1
\ifnum\count255<#1\repeat\rightarrow}}
\def\StretchLtArr#1{\,{\leftarrow\!\!\count255=0\loop\relbar
\joinrel\advance\count255 by1\ifnum\count255<#1\repeat}}

\def\StretchLRtArr#1{\,{\leftarrow\!\!\count255=0\loop\relbar\joinrel\advance
\count255 by1\ifnum\count255<#1\repeat\rightarrow\,\,}}

\def\mbox#1{{\leavevmode\hbox{#1}}}

\def\hspace#1{{\phantom{\mbox#1}}}
\def\oZ{\mbox{\open\char90}}

\def\al{\alpha}
\def\btau{{\bmit\tau}} 
\def\be{\beta}
\def\ga{\gamma}

\def\Ga{\Gamma}

\def\la{\lambda}

\def\si{\sigma}

\def\th{\theta}

\def\ze{\zeta}

\def\De{\Delta}

\def\caL{{\cal L}}

\def\caD{{\cal D}}

\def\zf{$\zeta$--function}
\def\zfs{$\zeta$--functions}
\def\hk{heat-kernel}


\def\frac#1/#2{\leavevmode\kern.1em
\raise.5ex\hbox{\the\scriptfont0 #1}\kern-.1em/\kern-.15em
\lower.25ex\hbox{\the\scriptfont0 #2}}
\def\sfrac#1/#2{\leavevmode\kern.1em
\raise.5ex\hbox{\the\scriptscriptfont0 #1}\kern-.1em/\kern-.15em
\lower.25ex\hbox{\the\scriptscriptfont0 #2}}

\def\gtorder{\mathrel{\raise.3ex\hbox{$>$}\mkern-14mu
             \lower0.6ex\hbox{$\sim$}}}
\def\ltorder{\mathrel{\raise.3ex\hbox{$<$}\mkern-14mu
             \lower0.6ex\hbox{$\sim$}}}

\def\semidirprod{\rlap{\ss C}\raise1pt\hbox{$\mkern.75mu\times$}}
\def\for{\lower6pt\hbox{$\Big|$}}
\def\fish{\kern-.25em{\phantom{abcde}\over \phantom{abcde}}\kern-.25em}


\def\boxit#1{\vbox{\hrule\hbox{\vrule\kern3pt
        \vbox{\kern3pt#1\kern3pt}\kern3pt\vrule}\hrule}}
\def\dalemb#1#2{{\vbox{\hrule height .#2pt
        \hbox{\vrule width.#2pt height#1pt \kern#1pt \vrule
                width.#2pt} \hrule height.#2pt}}}

\def\ol{\overline}
\def\frac#1#2{{{#1}\over{#2}}}


\def\cosec{{\rm cosec\,}}

\def\ie{{\it i.e. }}

\def\ket#1{\mid#1\rangle}
\def\br#1#2{\langle{#1}\mid{#2}\rangle}   

\def\Tr{{\rm Tr\,}}

\def\sumdash#1{{\mathop{{\sum}'}_{#1}}}

\def\Threej#1#2#3#4#5#6{\biggl({#1\atop#4}{#2\atop#5}{#3\atop#6}\biggr)}

\def\3j#1#2#3#4#5#6{\left\lgroup\matrix{#1&#2&#3\cr#4&#5&#6\cr}
\right\rgroup}

\def\man{{\cal M}}

\def\m?{\mgn{?}}

\def\beq{\begin{eqnarray}}
\def\eeq{\end{eqnarray}}


\def\aop#1#2#3{{\it Ann. Phys.} {\bf {#1}} ({#2}) #3}

\def\cmp#1#2#3{{\it Comm. Math. Phys.} {\bf {#1}} ({#2}) #3}
\def\cqg#1#2#3{{\it Class. Quant. Grav.} {\bf {#1}} ({#2}) #3}

\def\jmp#1#2#3{{\it J. Math. Phys.} {\bf {#1}} ({#2}) #3}
\def\jpa#1#2#3{{\it J. Phys.} {\bf A{#1}} ({#2}) #3}

\def\np#1#2#3{{\it Nucl. Phys.} {\bf B{#1}} ({#2}) #3}
\def\pl#1#2#3{{\it Phys. Lett.} {\bf {#1}} ({#2}) #3}

\def\prp#1#2#3{{\it Phys. Rep.} {\bf {#1}} ({#2}) #3}
\def\pr#1#2#3{{\it Phys. Rev.} {\bf {#1}} ({#2}) #3}

\def\prD#1#2#3{{\it Phys. Rev.} {\bf D{#1}} ({#2}) #3}

\def\rmp#1#2#3{{\it Rev. Mod. Phys.} {\bf {#1}} ({#2}) #3}

\def\zfp#1#2#3{{\it Z. f. Phys.} {\bf {#1}} ({#2}) #3}

\def\cras#1#2#3{{\it Comptes Rend. Acad. Sci. (Paris)} {\bf{#1}} (#2) #3}
\def\prs#1#2#3{{\it Proc. Roy. Soc.} {\bf A{#1}} ({#2}) #3}

\def\amsh#1#2#3{{\it Abh. Math. Sem. Ham.} {\bf {#1}} ({#2}) #3}
\def\am#1#2#3{{\it Acta Mathematica} {\bf {#1}} ({#2}) #3}
\def\aim#1#2#3{{\it Adv. in Math.} {\bf {#1}} ({#2}) #3}
\def\ajm#1#2#3{{\it Am. J. Math.} {\bf {#1}} ({#2}) #3}

\def\aom#1#2#3{{\it Ann. of Math.} {\bf {#1}} ({#2}) #3}
\def\cjm#1#2#3{{\it Can. J. Math.} {\bf {#1}} ({#2}) #3}

\def\dmj#1#2#3{{\it Duke Math. J.} {\bf {#1}} ({#2}) #3}
\def\invm#1#2#3{{\it Invent. Math.} {\bf {#1}} ({#2}) #3}

\def\jdg#1#2#3{{\it J. Diff. Geom.} {\bf {#1}} ({#2}) #3}

\def\jram#1#2#3{{\it J. f. reine u. Angew. Math.} {\bf {#1}} ({#2}) #3}
\def\jims#1#2#3{{\it J. Indian. Math. Soc.} {\bf {#1}} ({#2}) #3}
\def\jlms#1#2#3{{\it J. Lond. Math. Soc.} {\bf {#1}} ({#2}) #3}

\def\ma#1#2#3{{\it Math. Ann.} {\bf {#1}} ({#2}) #3}

\def\mz#1#2#3{{\it Math. Zeit.} {\bf {#1}} ({#2}) #3}
\def\ojm#1#2#3{{\it Osaka J.Math.} {\bf {#1}} ({#2}) #3}

\def\plms#1#2#3{{\it Proc. Lond. Math. Soc.} {\bf {#1}} ({#2}) #3}
\def\pgma#1#2#3{{\it Proc. Glasgow Math. Ass.} {\bf {#1}} ({#2}) #3}
\def\qjm#1#2#3{{\it Quart. J. Math.} {\bf {#1}} ({#2}) #3}

\def\rmjm#1#2#3{{\it Rocky Mountain J. Math.} {\bf {#1}} ({#2}) #3}

\def\tams#1#2#3{{\it Trans.Am.Math.Soc.} {\bf {#1}} ({#2}) #3}

\begin{title}
\vglue 1truein
\vskip15truept
\centertext {\Bigfonts \bf Spherical Casimir energies and} \vskip10truept
\centertext{\Bigfonts \bf Dedekind sums} \vskip 20truept
\centertext{J.S.Dowker\footnote{dowker@a35.ph.man.ac.uk}} \vskip 7truept
\centertext{\it Department of Theoretical Physics,} \centertext{\it The
University of
 Manchester,} \centertext{\it Manchester, England}
\vskip40truept
\begin{narrow}
Casimir energies on space--times having general lens spaces as their spatial
sections are shown to be given in terms of generalised Dedekind sums related
to Zagier's. These are evaluated explicitly in certain cases as functions of
the order of the lens space. An easily implemented recursion approach is
used.
\end{narrow}
\vskip 5truept
\vskip 60truept
\vfil
\end{title}
\pagenum=0
\newpage

\section{\bf 1. Introduction.}

Exact solutions have a fascination that transcends any physical realisation.
This is partly the reason for the use of toy, or cut down, models, the hope
being that the generality and elegance of the solution would make up for any
loss of realism. It is in this spirit that I have often resorted to
evaluations in the Einstein Universe, which, although being static, is not a
completely unrealistic cosmology, and it does form the instantaneous geometry
of a Friedman--Robertson--Walker space--time.

In some works, [\pref{DandB,Dow11}], I considered the Casimir effect on the
Einstein Universe, or three--sphere, looking at the factored case, S$^3/\Ga$,
and, in particular, at fixed--point free, homogeneous actions. Because of the
geometry of rotations, it is sufficient to evaluate for lens spaces, \ie for
$\Ga$ a cyclic group, $\oZ_q$, and combine these to get the other cases. This
has been done in [\pref{Dow11}] where attention was restricted to one--sided
(homogeneous) lens spaces, the simplest kind. In the present work, after a
question by Gregory Moore and David Kutasov, I wish to extend my treatment to
two--sided lens spaces the reason being that these are also exactly
computable, in one sense. The answers are no doubt academic and are presented
in accordance with the remarks above. However, there is current interest in
the possibility of a  topologically non--trivial Universe and my results may
be of relevance here. Relatedly, there has been recent discussion of the
implications of the S$^3/\Ga$ quotients for supersymmetry in the context of
the AdS/CFT correspondance and the brane world scenario [\pref{McInnes}].

Because the physical motivation is meant only to lead quickly to the
mathematical calculation, I concentrate on the total vacuum energy, as
opposed to the energy density. In the next section some basic facts are
outlined that lead as quickly as possible to the technicalities, which, I
admit, are my main interest in this work, which should be considered as a
continuation of [\pref{Dow11}].

One reason for the attraction of the three--sphere is that it is a group
manifold. Because of the isomorphism S$^3\sim$ SU(2), it is handy to
implement the action of $\Ga$ on the sphere as the combination of left and
right group actions on SU(2) \ie roughly, $\Ga\sim \Ga_L\times\Ga_R$. A
general point on the sphere S$^3$ is denoted by $q$, or, as an element of
SU(2), again by $q$ (or sometimes by $g$) and the general action $q\to\ga q$
is realised as $q\to\ga_Lq\ga_R$.

\section{\bf 2. Modes. Casimir energy.}
I intend to consider (massless) fields of spin 0, 1/2 and 1. The spin zero
expression for the Casimir energy has been given in [\pref{DandB}], and
repeated in [\pref{Dow11}], for double--sided actions. For one--sided
actions, the corresponding formulae for spins 1/2 and 1 have been detailed in
[\pref{Jadhav,Jadhav2}] and [\pref{DandJ}]. For completeness I wish to derive
the double--sided forms. The answers could be be deduced from the
off--diagonal \zf\ given in [\pref{Dow10}] but it is preferable to start
again. Unfortunately, a certain amount of dynamic and, worse, kinematic
recapitulation is required to reach the needed forms.

The equations of motion that concern me here are the Klein--Gordon, Dirac and
Maxwell equations. For mathematical convenience, I consider conformally
coupled scalars so that the eigenvalues of the relevant Helmholtz equation
are squares of integers.

In these cases there are no field theory divergences to worry about, in the
absence of fixed points, and a reasonably unambiguous formula for the Casimir
energy can be given as a regularised sum of the eigenvalues of the relevant
Hamiltonian operator which, for scalars, is the square--root of the Helmholtz
operator. The spectrum has positive and negative parts. In order to be more
specific, some mode properties are required and I now briefly repeat some
standard kinematics, [\pref{Dows,Dow10}], firstly on the full sphere.

For spinor fields (\ie any field with spin greater than zero), the Cartan
moving frame method projects everything onto one of the sets of Killing
fields on S$^3$. It is convenient to use the left--invariant set and we then
say we are using `right spinors'. These are scalars under left actions and,
further, the standard results of angular momentum theory can be used with no
changes in standard conventions.

For spin--zero, the angular momentum representation matrices,
$\caD^{L\,\, M}_N(g)$, $L=0,1/2,1,\ldots$, provide a complete set of modes
on the full three--sphere.

For higher spins, $j\,(=1/2,1,\ldots)$, spin--orbit coupling gives the modes
as the {\it spinor hyperspherical harmonics},
  $$\eqalign{
  Y^{j\,LM}_{mNJ}(q)&=\br{m,q}{JMLjN}\cr
  &=\bigg({(2L+1)(2J+1)\over2\pi^2a^3}\bigg)^{1/2}\caD^{LN'}_N(q)
  \Threej jLMm{N'}J .
  }
  \eql{hsh}
  $$

The equations of motion imply the massless polarisation conditions,
  $$\eqalign{
  J&=L\pm j\quad {\rm for}\,\,\,L\ge j\cr
  J&=L+ j\quad {\rm for}\,\,\,L< j\,,
  }
  $$
and then it can be shown by simple spin--orbit diagonalisation that the
eigenvalues of the Hamiltonian operators (the square--root of the Helmholtz
operator, the Dirac operator and the Maxwell curl) are,
  $$
  H\ket{JMLjN}=E^j_{L,J}\ket{JMLjN}\,,
  \eql{ham}
  $$
with
  $$
   E^j_{L,L\pm j}={1\over a}(j\pm\overline L),\quad \overline L=2L+1\,.
  $$
The energy spectrum on the full sphere is thus
  $$\eqalign{
  E^+_{\overline L}&={1\over a}(j+\overline L),\quad\overline L=1,2,\ldots\cr
  E^-_{\overline L}&={1\over a}(j-\overline L),\quad\overline
  L=2j+1,2j+2,\ldots\,.
  }
  \eql{spec}
  $$
I note that $E^+$ is positive and $E^-$ is negative, and that the spectrum on
the full sphere is symmetrical about zero.

As is well known, factoring the sphere to S$^3/\Ga$ modifies the spectrum by
eliminating certain modes and the spectrum is a subset of the full sphere
one. I denote the resulting degeneracies of the eigenvalues (\peq{spec}) by
$d^+(\overline L)$ and $d^-(\overline L)$ and define the handy positive and
negative Hamiltonian \zfs,
  $$\eqalign{
  \ze^+_j(s)&=a^s\sum_{\overline L=1}^\infty {d^+(\ol L)\over(\ol
  L+j) ^s}\cr
  \ze^-_j(s)&=a^s\sum_{\overline L=2j+1}^\infty {d^-(\ol L)\over(\ol
  L-j) ^s}=a^s\sum_{\overline L=1}^\infty {d^-(\ol L+2j)\over(\ol
  L+j) ^s}\,,\cr
  }
  \eql{zetaf}
  $$
in terms of which the regularised value of the Casimir energy is
  $$
  E_j=(-1)^{2j}{1\over4}h(j)\,\big(\ze^+_j(-1)+\ze^-_j(-1)\big)\,,
  \eql{casen}
  $$
where $h(j)$ is a spin degeneracy factor, $h(0)=1$ and $h(j)=2,\,j>0$.
Corners have been cut here in anticipation of the fact that there is no
divergence in the continuation of the \zfs\ at $-1$.

The problem now falls onto the evaluation of the degeneracies
\begin{ignore}
In general, the modes on the factored space, $\widetilde\man/\Ga$, can be
obtained by projection from those, $\widetilde\phi$, on the covering space,
$\widetilde \man$. This amounts to performing an average, or image sum,
followed, possibly, by a diagonalisation. The degeneracies can be obtained by
calculating the trace of this projection which yields, [\pref{Dow11}], the
standard formula
   $$
  d_\la={1\over|\Ga|}\sum_\ga\int_{\widetilde\man}\sum_i
  \widetilde\phi^*_{\la,i}(\ga^{-1} q)\,\widetilde
  \phi_{\la,i}(q)\,dq\,,
  \eql{degen1}
  $$
where $\la$ labels the eigenvalue and $i$ stands for the remaining labels. In
the present case where the $\widetilde\phi$ are given by the $Y$'s,
(\peq{hsh}), $\la=L$ and $i=(J,M,N)$.
\end{ignore}
which can be deduced in the following rather oblique way. Rather than analyse
the projected modes, it is clearer to consider the covariant \hk\ for, say
the relevant second order spin--$j$ propagation equation of which (\peq{hsh})
are eigenmodes and the eigenvalues are the squares of $E_{L,J}$, (\peq{ham}).
The associated degeneracies will be denoted by $d(L,J)$ and are related to
those in (\peq{zetaf}) by,
  $$
  d^+(\ol L)=d(L,L+j)\,,\quad d^-(\ol L)=d(L,L-j)\,.
  \eql{reldeg}
  $$

The spinor \hk\ on S$^3/\Ga$, ${\bf K}_\Ga(q,q')$, is given, as usual, in
terms of that, ${\bf K}(q,q')$, on S$^3$ by a pre--image sum. However,
because right spinors have a non--trivial behaviour under right actions, it
is necessary to provide compensating right `gauge' rotations at the
pre--image points to bring the beine into uniform alignment so that the
pre--image sum makes sense.

One is therefore interested in the traced \hk\ on the factored space,
  $$
  K_\Ga=\sum_{\ga_L,\ga_R}\int_{S^3/\Ga}dq\,\Tr\big[{\bf K}(q,\ga_L q\ga_R)
  \caD^j(\ga_R^{-1})\big]\,,
  \eql{k1}
  $$
where ${\bf K}$ is constructed from the eigenmodes (\peq{hsh}). In order to
use the properties of these, the integral in (\peq{k1}) is changed to one
over the full sphere by using the invariance of the volume, $dq$, under
symmetry actions and also the equation, following from the definition of
right spinors, [\pref{Dows}],
  $$
  {\bf K}(\xi q\eta,\xi q'\eta)=\caD^j(\eta^{-1}){\bf K}(q,q')\caD^j(\eta)\,.
  \eql{switch}
  $$
Then
  $$
  K_\Ga={1\over|\Ga|}\sum_{\ga_L,\ga_R}\int_{S^3}dq\,
  \Tr\big[{\bf K}(q,\ga_L q\ga_R)
  \caD^j(\ga_R^{-1})\big]\,.
  \eql{k2}
  $$

The switching equation, (\peq{switch}), shows that it is sufficient to know
${\bf K}(g)\equiv{\bf K}(g,{\rm id})$ because ${\bf K}(g',g)={\bf
K}(g^{-1}g',{\rm id})$ (note the order).

The construction of the \hk, ${\bf K}$, in terms of the modes, $Y$, gives us
the expression for the degeneracies ($|\man|=2\pi^2 a^3$),
  $$\eqalign{
 &d(L,J)=\cr &{\ol L\,\ol J\over|\Ga||\man|}\sum_{\ga_L,\ga_R}\Threej
 jLMmNJ\Threej{m'}{N'}JjLM\caD^{j\,\,m}_{m'}(\ga_R)
 \int_{S^3}dq\,\caD^{L\,\,N}_{N'}(\ga_Rq^{-1}\ga_Lq))\,. }
  $$
Group products and the orthogonality of the $\caD$'s easily yield
  $$
  \int_{S^3}dq\,\caD^{L\,\,N}_{N'}(\ga_Rq^{-1}\ga_Lq)={|\man|\over\ol L}
  \chi^{(L)}(\ga_L)\caD^{L\,\,N}_{N'}(\ga_R)
  $$
and then standard angular momentum manipulations in the end give simply,
  $$
   d(L,J)={1\over|\Ga|}\sum_{\ga_L,\ga_R}\chi^{(L)}(\ga_L)\,\chi^{(J)}(\ga_R)
   \eql{degen2}
  $$
in terms of the SU(2) characters,
  $$
  \chi^{(L)}(\ga)={\sin \ol L\th_\ga\over\sin\th_\ga}\,,
  $$
where $\th_\ga$ is the `radial' angular coodinate labelling the group
element, $\ga$.

For spins greater than zero the spectrum is no longer symmetrical about
zero, as is well known.
\section{\bf 3. The Casimir energy formulae.}

The equations (\peq{casen}), (\peq{zetaf}), (\peq{reldeg}) and (\peq{degen2})
can now be put together to give explicit summations for the Casimir energy.
Thus
  $$\eqalign{
  E_j={(-1)^{2j}\over4 |\Ga|}\,h(j)\lim_{s\to-1}a^s\sum_{\ga_L,\ga_R}
  \bigg[\sum_{n=1}^\infty&{1\over (n+j)^s}
  {\sin n\th_L\sin(n+2j)\th_R\over\sin\th_L\sin\th_R}+\cr
  &\sum_{n=2j+1}^\infty{1\over (n-j)^s}
  {\sin n\th_L\sin(n-2j)\th_R\over\sin\th_L\sin\th_R}\bigg]\,,
  }
  $$
where I have set $n=\ol L$ and $\th_L=\th_{\ga_L}$, $\th_R=\th_{\ga_R}$.

It must be admitted that this equation is meant to apply only for $j=0,1/2$
and $1$. The further evaluation is possible for all $j$ but the field
theoretic significance is unclear.

The only $n$-sum in the group sum to diverge as $s\to-1$ is that for
$\ga={\rm id}$, ($\th_R=\th_L=0$). So this term is separated and the limit
taken in the rest.

The identity contribution is
  $$
  E^{\rm id}_j={(-1)^{2j}\over4 |\Ga|}\,h(j)\lim_{s\to-1}a^s\bigg[\sum_{n=1}^\infty
  {n(n+2j)\over(n+j)^s}+\sum_{n=2j+1}^\infty
  {n(n-2j)\over(n-j)^s}\bigg]
  $$
which can be evaluated as Hurwitz \zfs\ giving, expanding the Bernoulli
polynomials, (see [\pref{Dow10}]),
  $$
  E^{\rm id}_j={(-1)^{2j}\over240a |\Ga|}\,h(j)\big(30j^4-20j^2+1\big)\,.
  \eql{iden}
  $$
Specific values are
  $$
  E^{\rm id}_0={1\over240a|\Ga|}\,,\quad E^{\rm id}_{1/2}={17\over960a|\Ga|}\,,
  \quad E^{\rm id}_1={11\over120a|\Ga|}\,,
  \eql{idens}
  $$
which, apart from the volume factor, $1/|\Ga|$,  are the known full--sphere
values.

The harder part of the calculation is the evaluation of the non--identity
contribution,
  $$
  \sumdash{\ga_L,\ga_R}
  \bigg[\sum_{n=1}^\infty(n+j)
  {\sin n\th_L\sin(n+2j)\th_R\over\sin\th_L\sin\th_R}+
  \sum_{n=2j+1}^\infty(n-j)
  {\sin n\th_L\sin(n-2j)\th_R\over\sin\th_L\sin\th_R}\bigg]\,.
  \eql{nic}
  $$

It turns out that the integer and half--odd--integer $j$ cases differ and so
I treat them separately. For integer $j$, a shift of summation variable
yields the more symmetrical form,
  $$
  \sumdash{\ga_L,\ga_R}{1\over\sin\th_L\sin\th_R}
  \sum_{n=j}^\infty n\bigg[\sin (n-j)\th_L\sin(n+j)\th_R+
  \sin (n+j)\th_L\sin(n-j)\th_R\bigg]\,.
  $$
and simple trigonometry produces,
  $$
  \sumdash{\ga_L,\ga_R}{1\over\sin\th_L\sin\th_R}
  \sum_{n=j}^\infty n\big(\cos n\be\cos j\al-\cos n\al\cos j\be\big)\,.
  $$
where
  $$
  \al=\th_R+\th_L\,,\quad \be=\th_R-\th_L\,.
  $$
The sum over $n$ can be done,
  $$
-{1\over2}\sumdash{\al,\be}\bigg[{\cosec^2\be/2\cos j\al-\cosec^2\al/2\cos
j\be\over \cos\al-\cos\be}\bigg]\,.
  $$

While this expression can be taken further for any $j$, I now set just $j=0$
and $j=1$ for simplicity, when again simple algebra gives
  $$-{1\over2}\sumdash{\al,\be}\bigg[{\cosec^2\be/2-\cosec^2\al/2
  \over \cos\al-\cos\be}\bigg]=
 -{1\over4}\sumdash{\al,\be}\cosec^2\be/2\,\,\cosec^2\al/2 \,,
 \eql{sp0}
  $$
for $j=0$ and
  $$\eqalign{
 -{1\over2}\sumdash{\al,\be}&\bigg[{\cosec^2\be/2\cos \al-\cosec^2\al/2\cos
 \be\over \cos\al-\cos\be}\bigg]=\cr
 -&{1\over2}\sumdash{\al,\be}\big(\cosec^2\al/2+\cosec^2\be/2-{1\over2}
 \cosec^2\al/2\,\,\cosec^2\be/2\big)\,,
 }
 \eql{sp1}
  $$
for $j=1$.

For spin--half I return to (\peq{nic}) and rewrite the sum as one over odd
integers, $\ol n=2n+1$. Trigonometry and use of the $\al$ and $\be$ now give
  $$
  \sumdash{\al,\be}{1\over\cos\al-\cos\be}
  \sum_{\ol n=1,3}^\infty \ol n\,\big(\cos \ol n\be/2\cos \al/2
  -\cos \ol n\al/2\cos \be/2\big)\,,
  $$
and the $\ol n$ summation is done using
  $$
    \sum_{\ol n=1,3}^\infty\sin\ol n\th={1\over2}\cosec\th\,,\quad
   \sum_{\ol n=1,3}^\infty \ol n\,\cos \ol
   n\th=-{1\over2}\cot\th\,\cosec\th\,,
  $$
which yields,
  $$
  {1\over8}\sumdash{\al,\be}\cot\al/2\,\cosec\al/2\,\,
  \cot\be/2\,\cosec\be/2\,.
  \eql{sphalf}
  $$
\section{\bf 4. The Casimir energy calculated on lens spaces.}

Reinstating the factors, the expressions for the three Casimir energies are
  $$\eqalign{
  E_0&={1\over a|\Ga|}\bigg[{1\over240}
  -{1\over16}\sumdash{\al,\be}\cosec^2\be/2\,\,\cosec^2\al/2\bigg] \cr
  E_{1/2}&={1\over a|\Ga|}\bigg[{17\over960}
  +{1\over8}\sumdash{\al,\be}\cot\al/2\,\cosec\al/2\,\,
  \cot\be/2\,\cosec\be/2\bigg]
  \cr
  E_1&={1\over a|\Ga|}\bigg[{11\over120}
  +{1\over4}\sumdash{\al,\be}\big(\cosec^2\al/2+\cosec^2\be/2-{1\over2}
 \cosec^2\al/2\,\,\cosec^2\be/2\big)\bigg]\,.\cr
  }
  \eql{vens}
  $$

The expression for $E_0$ was given in [\pref{DandB}]. For one--sided actions
(\ie $\al=\pm\be$), the expressions reduce to the ones in
[\pref{Dow11,DandJ}].

For the lens space, $L(q;l_1,l_2)$, the angles $\al$ and $\be$ take the
values
  $$
  \al={2\pi p\nu_1\over q}\,,\quad \be={2\pi p\nu_2\over q}\,,
  \eql{angles1}
  $$
where $p,\,=1,\ldots,q-1\,,$ labels $\ga$. $\nu_1$ and $\nu_2$ are integers
coprime to $q$, with $l_1$ and $l_2$ their mod $q$ inverses. The simple,
`one--sided' lens space, $L(q;1,1)$, corresponds to setting
$\nu_2=\nu_1=\nu=1$, say, so that $\th_L=0$ and $\th_R=2\pi p/q$. These were
discussed in [\pref{Dow11}]. I now examine the general case and begin with
the integer spin forms which involve only powers of cosecants. The single
sums of cosec$^2$ are classic,
  $$
  S(q;\nu)={1\over q}\sum_{p=1}^{q-1} \cosec^2{\pi p\nu\over q}={q^2-1\over3q},
  \eql{csc2}
  $$
and independent of $\nu$, being sums over the roots of unity ordered in
different ways as in the stellated polygon.

I require the harder, split summations,
  $$
  S(q;\nu_1,\nu_2)={1\over q}\sum_{p=1}^{q-1} \cosec^2{\pi p\nu_1\over q}\,
  \cosec^2{\pi p\nu_2\over q}
  \eql{csc4}
  $$
where $\nu_1$ and $\nu_2$ are fixed integers from 1 to $q-1$. For these
particular forms I can fortunately make use of the computations of Zagier
[\pref{Zagier}], who has defined a generalised Dedekind sum (I change his
notation slightly)
   $$
   \tau(q;\nu_1,\nu_2,\ldots,\nu_n)=(-1)^{n/2}{1\over q}\sum_{p=1}^{q-1}
   \cot{\pi p\nu_1\over q}  \ldots \cot{\pi p\nu_n\over q}\,.
   \eql{zagded}
  $$
Here $q$ is a positive integer, $\nu_1,\ldots,\nu_n$ are integers prime to
$q$ and $n$ is even. I note the important fact that one of the sets of roots
of unity can be reordered to the canonical one while maintaining the relation
between all the sets so keeping the sum the same. Formally this is done by
multiplying all the $\nu_i$ by the mod $q$ inverse, $l_1$, say.

I need the four--dimensional case, $\tau(q;\nu_1,\nu_1,\nu_2,\nu_2)$, and, as
said, one can put $\nu_1=1$ without loss of generality. So I look at
$\tau(g;1,1,\nu,\nu)$ and note the simple relation that takes me back to the
cosecant sums,
  $$\eqalign{
  S(q;\nu,1)&=\tau(q;1,1,\nu,\nu)+2S(q;1)-{q-1\over q}\cr
  &=\tau(q;1,1,\nu,\nu)+{(q-1)(2q-1)\over3q}\,,
  }
  \eql{reln}
  $$
$S(q;1)$ being given by (\peq{csc2}).

In his Table 3, Zagier has listed some four--dimensional examples, of which
$d(q;1,1,3,3)$ and $d(q;1,1,4,4)$ are most relevant for us. The first
quantity has also been evaluated by Harvey {\it et al}, [\pref{HKMM}], in
its cosec version. Employing Zagier's expression, one finds,
  $$
  S(q;3,1)={q^4+210q^2\pm80q-291\over405q}\,,\quad q=\pm1\,\,({\rm mod}\, 3)
  $$
in agreement with [\pref{HKMM}]. Note that use of the cosec form has resulted
in a more symmetrical result.

My intention here is to take the computations further for
$\tau(q;1,1,\nu,\nu)$ with larger $\nu$.
\section{\bf 5. Evaluation of the Dedekind sums.}

The essential calculational tool is the $(n+1)$--term reciprocity law obeyed
by the Dedekind sums (\peq{zagded}) (I state it for the 4--dimensional case),
  $$
  \sum_{j=0}^4 \tau(\nu_j;\nu_0,\ldots,\widehat \nu_j,\ldots,\nu_4)=
  \phi_4(\nu_0,\ldots,\nu_4)\,,
  \eql{recip}
  $$
where $\nu_0,\ldots,\nu_4$ are pairwise coprime positive integers, the hat
signifies omission of the indicated term and $\phi_4$ is a function whose
form can be calculated.

Zagier derives the relation (\peq{recip}) using contour integration which
indicates why the construction of $\tau$ in terms of products of cotangents
is so convenient as the pole defining function is another cotangent, giving a
symmetrical structure. This leads to the expression for $\phi_n$,
  $$
  \phi_n(\nu_0,\ldots,\nu_n)=1-{\caL_n(\nu_0,\ldots,\nu_n)
  \over \nu_0\ldots\nu_n}
  $$
where the polynomial, $\caL_n(\nu_0,\ldots,\nu_n)$ is the coefficient of
$t^n$ in the power series expansion of
  $$
  \prod_{j=0}^n{\nu_jt\over\tanh\nu_jt}\,.
  $$
The individual expansions of the $\coth$s give an explicit multiple sum
expression for this quantity (Berndt, [\pref{Berndt7}]) which is better
expressed as a generalised Bernoulli polynomial but I do not need this
cosmetically compact form here.

The relation (\peq{recip}) cannot be applied immediately to the case here
because the members of the set, $(q,1,1,\nu,\nu)$, are not mutually coprime.
However $\nu$ is prime to $q$ so, following Zagier, I make use of periodicity
to replace $\tau(q;1,1,\nu,\nu)$ by $-\tau(q;1,1,\nu,q-\nu)$ so that the set
$(q,1,1,\nu,q-\nu))$ is coprime and (\peq{recip}) can be applied. In writing
(\peq{recip}) out I immediately use the fact that
$\tau(1;\nu_1,\nu_2,\nu_3,\nu_4)$ is always zero which produces a recursion
formula, as is now shown. The reciprocal relation reduces directly to
  $$
   \tau(q;1,1,\nu,q-\nu)+\tau(\nu;1,1,q,q-\nu)+\tau(q-\nu;1,1,\nu,q)
   =\phi_4(q,1,1,\nu,q-\nu)
   \eql{recip2}
  $$
For short, set
  $$
  \tau_q=\tau(q;1,1,\nu,\nu)
  $$
so the first term on the left in (\peq{recip2}) is $-\tau_q$ and the third
term is $\tau_{q-\nu}$ as can be shown using mod $q-\nu$ periodicity. The
second term is to be taken to the right--hand side and together with $\phi_4$
will form the driving term of the recursion which, as is seen, successively
reduces $q$ by $\nu$. The driving term has to be explicit. $\phi_4$ is given
by Zagier and equals, in this case,
  $$
  \phi_4(q,1,1,\nu,q-\nu)=1-
  {3\nu^4-6\nu^3q+9\nu^2q^2+20\nu^2-\nu q^3-20\nu q+3q^4+20q^2+3
  \over45\nu q(q-\nu)}
  \eql{phi4}
  $$

A little more argument is needed for $\tau(\nu;1,1,q,q-\nu)$. It is necessary
to break up the solution according to the residue classes mod $\nu$ of $q$.
Thus, if $q=n\nu+x$, $\tau(\nu;1,1,q,q-\nu)=\tau(\nu;1,1,x,x)$ and this
`constant' has to be calculated separately for each $x$, ($1\le x\le \nu$),
for the given $\nu$.

As an introduction to the general method, I check the case mentioned earlier,
$\nu=3$. There are two ranges for $q$, \ie $q=\pm1$ mod $3$. One therefore
requires $\tau(3;1,1,1,1)$ and $\tau(3;1,1,-1,-1)$, $=\tau(3;1,1,1,1)$ also.
This quantity follows from the classic summation,
  $$
  \tau(q;1,1,1,1)={1\over q}\sum_{p=1}^q\cot^4{\pi p\over q}
  ={(q-1)(q-2)(q^2+3q-13)\over45q}\,,
  \eql{cot4}
  $$
as $2/27$. Hence, evaluating the function $\phi_4$, (\peq{phi4}), the
recursion reads,
  $$
  \tau_q+{142\over135q}=
  \tau_{q-3}+{142\over135(q-3)}+
  {q^2\over45}-{q\over15}-{17\over45}\,,
  $$
in both cases. This can be iterated down to $\tau_1=0$ for $q=1$ mod $3$ and
to $\tau_2=\tau(2;1,1,3,3)=\tau(2;1,1,1,1)=0$, also, for $q=-1$ mod $3$. Thus
for $q=3[q/3]+1$,
  $$\eqalign{
  \tau_q&=-{142\over135q}+\sum_{k=1}^{[q/3]}\bigg({(3k+1)^2\over45}
  -{3k+1\over15}-{17\over45}\bigg)+{142\over135}\cr
  &={(q-1)(q^3+q^2-59q+426)\over 405q}\,,\cr
  }
  $$
and for $q=3[q/3]+2$,
  $$\eqalign{
  \tau_q&=-{142\over135q}+\sum_{k=1}^{[q/3]}\bigg({(3k+2)^2\over45}
  -{3k+2\over15}-{17\over45}\bigg) +{142\over270}\cr
  &={(q-2)(q^3+2q^2-56q+213)\over405q}\,,
  }
  \eql{zag2}
  $$
which are the forms listed in Zagier, [\pref{Zagier}].
\section{\bf 6. Automatic evaluation.}

The above `hand' calculation is typical. I wish to see how far it can be
automated for any, but fixed, $\nu$.

The form of the function $\phi_4$ allows one to extract the combination,
  $$
  \si_q=\tau_q +{3\nu^4+20\nu^2+3\over45\nu^2q}\,,
  \eql{sig}
  $$
and to write the general recursion as,
  $$
  \si_q-\si_{q-\nu}=\tau(\nu;1,1,x,x)-1+{3q^2-3\nu q+6\nu^2+20\over45\nu}\,.
  $$
For $q=[q/\nu]\nu+x$, for given $x$, this is taken down to $\si_x$ to give
\footnote{ $x$ does not alter during the iteration.}
  $$\eqalign{
  \si_q&=\si_x+\sum_{k=1}^{[q/\nu]}\bigg[\tau(\nu;1,1,x,x)-1+{3(\nu k+x)^2-
  3\nu(\nu k+x)+6\nu^2+20\over45\nu}\bigg]\cr
  &=\si_x+\big(\tau(\nu;1,1,x,x)-1\big){(q-x)\over\nu}
  +{(q-x)(q^2+qx+x^2+5\nu^2+20)\over45\nu^2}\,.
  }
  \eql{iter}
  $$
Everything is therefore explicit except the starting value, $\si_x$, and the
`constant', $\tau(\nu;1,1,x,x)$ which have to be found separately. The tables
in Zagier could be consulted after suitable transformations using the basic
properties of the $\tau$'s to reduce the arguments to those appearing in the
table but I here continue with the recursion approach
for a self--contained treatment.

Some streamlining of notation is indicated. I set,
  $$\eqalign{
  \tau_q&\equiv\tau(q;1,1,\nu,\nu)=\tau(q,\nu)\,,\cr
  }
  $$
and rewrite the iteration result (\peq{iter}) in terms of the $\tau$'s,
$$\eqalign{
  \tau(q,\nu)=\tau(x,\nu)+{(q-x)\over\nu}\,\tau(\nu,x)
  +F(q,\nu,x)\,,
  }
  \eql{iter5}
  $$
where,
  $$\eqalign{
  F(a,b,c)={1\over45 b^2ac}\bigg(a(&3b^4+20b^2+3)
  -ac^4-5a c^2(b^2-9b+4)-\cr
  &c\big(3b^4+5b^2(4-a^2)+45b  a^2-a^4-20a^2+3\big)\bigg)\,.
  }
  \eql{eff}
  $$
I recall that $q$ is arbitrary, $\nu$ is a fixed number, prime to $q$ and
 $$
 x=q\,{\rm mod}\, \nu
 \eql{red2}
 $$
is to be considered a chosen number, also prime to $q$.

In the first term on the right--hand side of (\peq{iter5}), $\nu$ can be
reduced mod $x$, defining,
  $$
  x'=\nu\,\, {\rm mod}\, \,x\,,
  \eql{red}
  $$
($x'<x<\nu<q$). Hence follows the fundamental relation,
  $$\eqalign{
  \tau(q,\nu)=\tau(x,x')+{(q-x)\over\nu}\,\tau(\nu,x)
  +F(q,\nu,x)\,.
  }
  \eql{iter2}
  $$

I introduce the series of Euclidean relations,
  $$
  x_i=x_{i-2}\,{\rm mod}\,x_{i-1}\,,\quad x_{i-1}<x_i\,,
  \eql{eucl}
  $$
of which (\peq{red2}) and (\peq{red}) are the first two examples, with
$x_0=q$, $x_1=\nu$, $x_2=x$ and $x_3=x'$. The recursion (\peq{iter2}) then
becomes, with a more systematic labelling,
  $$\eqalign{
  \tau(x_i,x_{i+1})={(x_i-x_{i+2})\over x_{i+1}}\,\tau(x_{i+1},x_{i+2})
  +\tau(x_{i+2},x_{i+3})+F(x_i,x_{i+1},x_{i+2})\,,
  }
  \eql{iter3}
  $$
which begins with $i=0$ giving (\peq{iter2}). The task, to repeat, is to find
the `constants' $\tau(x_1,x_2)$ and $\tau(x_2,x_3)$ by iterating
(\peq{iter3}).

The algorithm (\peq{eucl}) ceases when one of the $x_i$ reaches 1. Assume
that this happens when $i=t$ ($t$ is usually small). The iteration,
(\peq{iter3}), will also come to an end when $i+2=t$ because $\tau$ is known
whenever one of the arguments is 1,
  $$\eqalign{
  \tau(b,1)&={(b-1)(b-2)(b^2+3b-13)\over45b}\cr
  \tau(1,d)&=0\,.
  }
  \eql{vals}
  $$
This means that $\tau(x_{t-2},x_{t-1})$ is known from the recursion which can
be rolled up from the bottom by successive back substitution. I find, for
example, the first non--trivial value,
  $$\eqalign{
   \tau(a,b)&={a-1\over b}{(b-1)(b-2)(b^2+3b-13)\over45b}+F(a,b,1)\cr
   &={(a-1)\big(a^3+a^2+a(b^4-15b^2-5)+3b^4+20b^2+3\big)\over45 ab^2}\,,
   }
   \eql{nt}
  $$
where $(a,b,1)$ are the last three numbers in the Euclidean algorithm,
$(x_{t-2},x_{t-1},x_t)$. In particular, if $a=b+1$,
  $$
  \tau(b+1,b)={b(b-1)(b^2+5b-9)\over45(b+1)}\,,
  $$
which also follows simply from the reduction,
  $$\eqalign{
  \tau(b+1;1,1,b,b)&=\tau(b+1;-b,-b,b,b)\cr
  &=\tau(b+1;-1,-1,1,1)=\tau(b+1;1,1,1,1)\,,
  }
  $$
and (\peq{cot4}).

Likewise, but more generally, one finds (see also the Appendix),
  $$\eqalign{
  \tau(nb+1;1,1,n,n)=
  {nb\big(nb^3+4b^2+nb(n^2-15)+4n^2+5\big)\over45(nb+1)}\,.
  }
  \eql{exeq1}
  $$

Similar reductions have to be performed in order to reach one of the given
forms in Zagier's table. My formula has the manipulations built in.

\begin{ignore}
The systematic elimination is performed by matrix inversion and a more
compressed notation is useful. I define,
  $$\eqalign{
   a_i={x_i-x_{i+2}\over x_{i+1}}\cr
   \tau_i=\tau(x_i,x_{i+1})\cr
   f_i=F(x_i,x_{i+1},x_{i+2})\,,
   }
  $$
so that the recursion (\peq{iter3}) becomes
  $$
  \tau_i-a_i\tau_{i+1}-\tau_{i+2}=f_i\,,
  \eql{iter4}
  $$
where, in general, $i$ runs from 1 to $(t-2)$m with $\tau_t=0$. Writing
(\peq{iter4}) out in columns, the last two columns can be dropped and their
contents, (from the bottom two rows), which are known from (\peq{vals}),
transferred to the right--hand side. This produces  a $(t-2)\times(t-2)$
matrix inversion problem ${\bf A}\btau={\bf g}$ for the vector
$\btau=(\tau_1,\ldots,\tau_{t-2})$, with
  $$
  {\bf A}=\left(\matrix{1&-a_1&-1&0&0&\ldots\cr
                        0&1&-a_2&-1&0&\ldots\cr
                        0&0&1&-a_3&-1&\ldots\cr
                        \vdots&\vdots&\vdots&\vdots&\vdots\cr
  }\right)
  $$
and
  $$
  {\bf g}=\left(\matrix{f_1\cr f_2\cr f_3\cr\vdots\cr g_{t-3}\cr g_{t-2}\,,
  }\right)
  $$
where
  $$\eqalign{
  g_{t-3}=f_{t-3}+\tau_{t-1}\cr
  g_{t-2}=f_{t-2}+a_{t-2}\tau_{t-1}\,,
  }
  $$
which are known.

The inverse ${\bf A}^{-1}$ can be expressed in terms of the determinants
  $$
  \De_n=\left|\matrix{ -a_1&-1&0&0&0&\ldots&0&0\cr
                       1&-a_2&-1&0&0&\ldots&0&0\cr
                       0&1&-a_3&-1&0&\ldots&0&0\cr
                       \vdots&\vdots&\vdots&\vdots&\vdots&&\vdots&\vdots\cr
                       0&0&0&0&0&\dots&1&-a_n
  }\right|
  $$

In practice, it is sufficient to calculate only the first component of $\btau$
  $$
  \tau_1=
  $$
\end{ignore}

The systematic elimination is therefore performed by inverse recursion and a
more compressed notation is useful. I define
   $$\eqalign{
   a_k&={x_{t+1-k}-x_{t+3-k}\over x_{t+2-k}}\cr
   \tau_k&=\tau(x_{t+1-k},x_{t+2-k})\cr
   f_k&=F(x_{t+1-k},x_{t+2-k},x_{t+3-k})\,,
   }
  $$
so that the two--term recursion (\peq{iter3}) becomes
  $$
  \tau_k=a_k\tau_{k-1}+\tau_{k-2}+f_k\,,\quad k=3\ldots t\,,
  \eql{iter4}
  $$
with the starting values,
  $$\eqalign{
   \tau_1&=0\cr
   \tau_2&={(b-1)(b-2)(b^2+3b-13)\over45b}\,,\quad b=x_{t-1}\,.
   }
  $$

To make everything plain, my two step procedure is to compute the quantity I
want $\tau(q,\nu)=\tau(x_0,x_1)$ from (\peq{iter3}) at $i=0$ by finding the
constants $\tau(x_1,x_2)$ and $\tau(x_2,x_3)$ from the recursion
(\peq{iter4}), which has to be taken to $k=t-1$ and $k=t$. The reason is that
this second stage is purely numerical. The variable $q$ is arbitrary, $\nu$
is a constant and $x$ is a number chosen in the range $1\le x< \nu$. (It may
be possible to leave $x$ variable, but I prefer to set it by hand.)

The first act is the standard computation of the Euclidean algorithm
(\peq{eucl}) for given $x_1=\nu$ and $x_2=x$. It is then straightforward to
implement the recursion to produce the polynomials in $q$ in one go. I coded
the entire process quite compactly in DERIVE.

\section {\bf 7. Some results.}
The general expression for the cosecant sum, $q\,S(q;\nu,1)$, is a polynomial
which I write in the following way,
  $$
   q\, S(q;\nu,1)= A\big(q^4+Bq^2+Cq+D\big)\,,
   \eql{poly4}
  $$
where, introducing $x=q$ mod $\nu$, the coefficients $A,B,C,D$ depend on
$\nu$ and $x$. $A,B,D$ are unchanged under $x\to\nu-x$, while $C$ changes to
sign. So $x$ can be restricted to $1\le x\le \nu/2$, and, of course, $x$ and
$\nu$ are coprime.

When $x=1$ or 2, there are relations among the coefficients because the {\it
cotangent} polynomials vanish when $q=1$ or $2$, respectively.

The factor of $A$ has been extracted because it is known from the obvious
$q\to\infty$ limit of (\peq{csc4}),\footnote{ This can be applied also to the
general sum, (\peq{zagded}).}
  $$
  A=2{\ze_R(4)\over\pi^4\nu^2}={1\over45\nu^2}\,.
  \eql{Ay}
  $$
(The factor of 2 is a consequence of the periodicity of the summand.) Hence
only $B,C$ and $D$ need be displayed which I do in the following table for a
few low values of $\nu$ and $x$,
  $$
  \matrix{\nu&x&B&C&D&&&\nu&x&B&C&D\cr
           3&1& 210&80&-291&&&9&1&7770&12320&-20091\cr
           4&1& 490&360&-851&&&9&2&3450&4960&-20091\cr
           5&1& 994&1008&-2003&&&9&4&3450&-640&-20091\cr
           5&2& 706&144&-2003&&&10&1&11494&19008&-30503\cr
           6&1& 1830&2240&-4071&&&10&3&4006&3456&-30503\cr
           7&1& 3130&4320&-7451&&&11&1&16450&28080&-44531\cr
           7&2& 1690&1440&-7451&&&11&2&6370&12240&-44531\cr
           7&3& 1690&0&-7451&&&11&3&4930&6480&-44531\cr
           8&1& 5050&7560&-12611&&&11&4&4930&3600&-44531\cr
           8&3& 2170&1080&-12611&&&11&5&6370&-2160&-44531\cr
          }
  $$

The lowest case, $\nu=2$, for which $B=70,\,C=0$ and $D=71$  can be obtained
directly using the relation $\cosec^22\th=(\cosec^2\th+\sec^2\th)/4$ and the
classic, single summations of $\cosec^4$, $\cosec^2$ and $\tan^2$.

\section{\bf 8. Some Casimir values.}

In this paper I consider only the integer spins, 0 and 1. The basic
expressions for the Casimir energies are given in (\peq{vens}).

In terms of the sums discussed above, the spin 0 formula on the lens space
$L(q;1,l)$, where $l$ is the mod $q$ inverse of $\nu$, is,
  $$
  E_0(q,\nu)={1\over a}\bigg[{1\over 240q}-{1\over16}S(q;\nu,1)\bigg]\,,
  \eql{0en}
  $$
and for spin--1,
  $$
  E_1(q,\nu)={1\over 2a}\bigg[{11\over
  60q}+S(q;1)-{1\over4}S(q;\nu,1)\bigg]\,.
  \eql{1en}
  $$
\vskip5truept As a sample, I plot in fig.1 the spin--0 energy as a function
of the order, $q$, for a fixed twisting, $\nu=5$,

\input epsf
\epsfbox{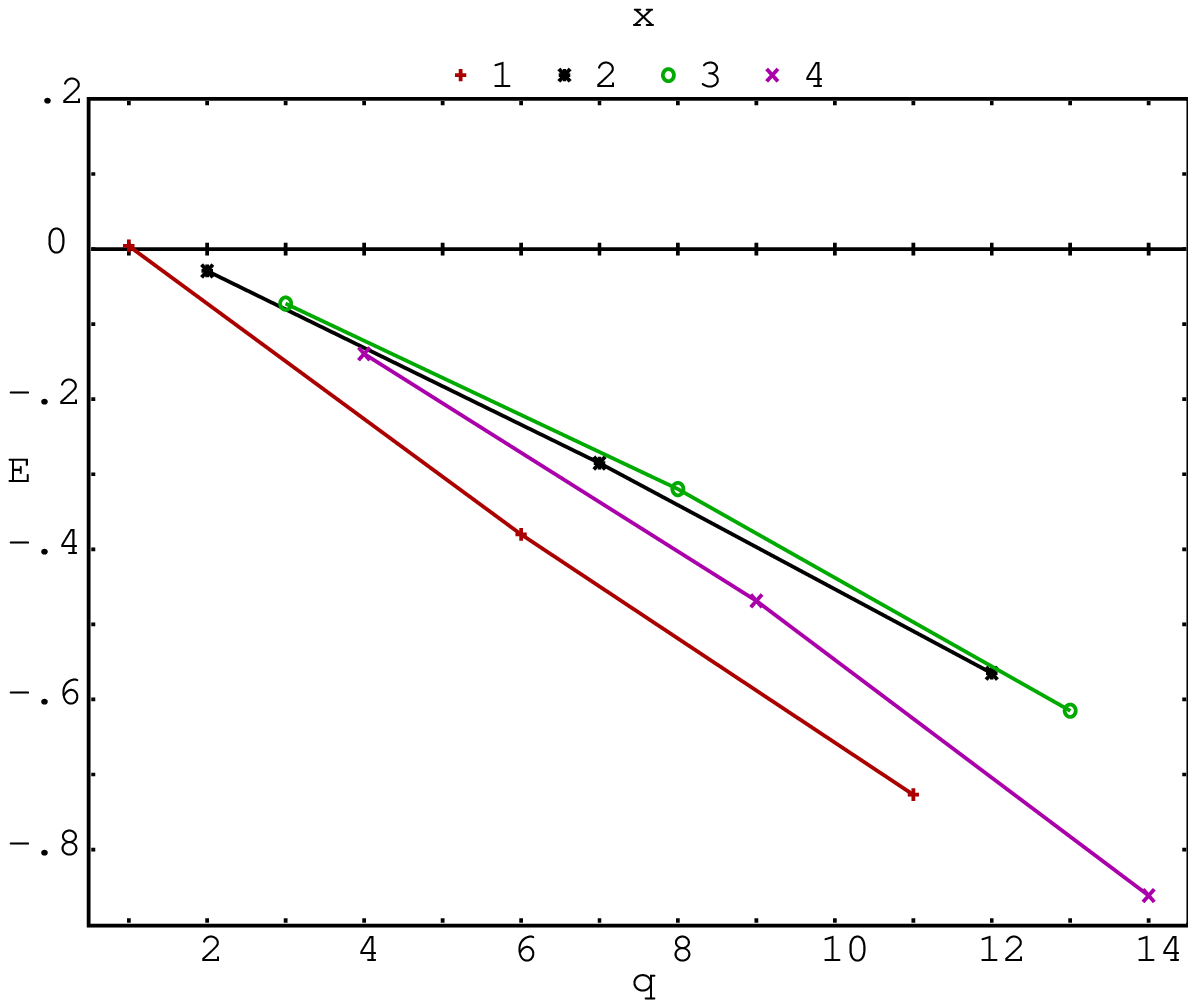} \vskip5truept {fig1. Casimir energy E for conformal
scalars on lens spaces of order q and $\hspace{******}\,\,\,$  twisting
$\nu=5$. $x=q\,\,{\rm mod}\,\,5$} \vskip 30truept Fig.2 plots, typically,
the Casimir energy against the twisting $\nu$ for a fixed order $q=29$. It
exhibits the symmetry about the value $\nu=q/2$ and a series of minima the
significance of which I do not know. The spin--one graph is similar except
it goes positive for certain values of $\nu$.

\vskip 25truept

\input epsf
\epsfbox{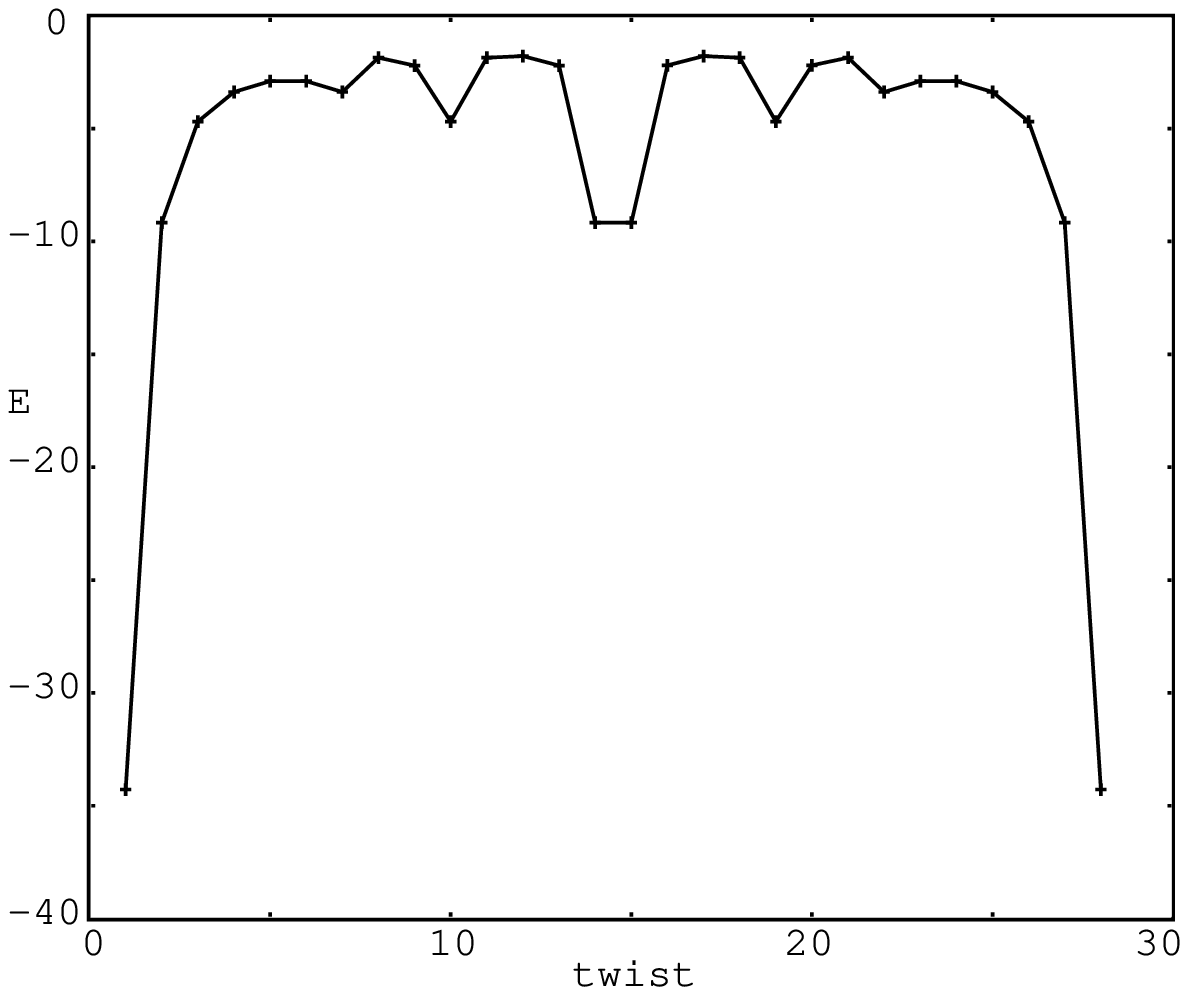} \vskip10truept {fig2. Casimir energy E for conformal
scalars on a lens space of order 29 for $\hspace{******}\,\,\,$  twistings
$\nu=1$ to 28}.

\newpage

\section{\bf 9. Conclusion and discussion.}

The values for the Casimir energies presented here, and in our earlier works,
are doubtless unpractical but their exact determination is not without
methodological value.

Harvey {\it et al}, [\pref{HKMM}], present an algorithm for the computation
of the cosecant sums, which can also be automated. In content, it is, of
course, the same as that given here. However I feel the present approach via
the cotangent sums is preferable as being more explicit and systematic.

There exist many generalisations of the original Dedekind sum. Some of the
older ones are covered in the book by Rademacher and Grosswald,
[\pref{RandG}]. A more recent sum, extending that of Zagier, has been
proposed by Beck, [\pref{Beck}], which involves replacing each cotangent in
(\peq{zagded}) by a higher derivative. Beck employs contours to determine the
reciprocity relation which could be used to compute the cosecant sums
directly by simply restricting to first derivatives. I have not checked this.
The article by Berndt and Yeap, [\pref{BandY}], contains further references.

Unfortunately I could not find the spin--half summations, (\peq{sphalf}), so
their reciprocal relation will have to be sought using contours, {\it ab
initio}.

Also, as a challenge, if one could evaluate the even higher expressions,
  $$
  \tau(q;1,1,\nu,\nu,\ldots,\mu,\mu)={1\over q}\sum_{p=1}^{q-1}
  \cot^{2}{\pi p\over q}\,\cot^{2}{\pi p\nu\over q}
  \ldots\cot^2{\pi p\mu\over q}\,,
  \eql{cotn}
  $$
 the cosecant sums,
   $$
   {1\over q}\sum_{p=1}^{q-1}
  \cosec^{2}{\pi p\over q}\,\cosec^{2}{\pi p\nu'\over q}
  \ldots\cosec^2{\pi p\mu'\over q}\,,\quad (\nu',\ldots,\mu')\subset
  (\nu,\ldots,\mu)\,,
  \eql{cscn}
  $$
would follow on binomial expansion. Such sums would be needed when evaluating
Casimir energies on the higher spheres.

For the case here, a more careful independent analysis of the asymptotic
behaviour of the cosecant, or cotangent sums as $q\to\infty$ would yield
general information about the coefficients in (\peq{poly4}).
\section {\bf Appendix.}

I make some remarks on the evaluation of the Dedekind sums not necessary for
the immediate prosecution of the calculation of the Casimir energies but
which extend the results a little.

In the preceding calculation the Euclidean algorithm was taken down to its
$x_t=1$ limit, which always works. However it is possible to stop if the
penultimate value, $x_{t-1}$, equals 2 because $\tau$ is also known when one
argument is 2,
  $$\eqalign{
  \tau(x_{t-1},x_t)&=\tau(2,1)=0\cr
  \tau(a,2)&={(a-1)(a^3+a^2-49a+131)\over180a}\,.
  }
  $$
The second expression checks against the result of the earlier recursion,
and is in Zagier's list, but the point is that it follows {\it directly}
from the standard sums of (even) powers of cotangents (or of cosecants)
after using the simple trigonometric formula,
  $$
  2\cot\th\,\cot2\th=\cot^2\th-1\,.
  \eql{cotid}
  $$

For convenience, I denote the last four entries in the Euclidean algorithm,
$(x_{t-3},x_{t-2},x_{t-1}, x_t)$, by $(d,a,b,c)$ with $b=2, c=1$. The
algorithm implies that $a$ is odd, $a=2n+1$, and also that $d=ma+b$,
($n,n\in\oZ$).

The first non--trivial result of the backwards recursion is
  $$
  \tau(d,a)={d-b\over a}\, \tau(a,b)+F(d,a,b)\,,
  $$
or, putting in the values,
  $$\eqalign{
  \tau\big(m(2n+1)+2,\,&2n+1\big)\cr
  =&\,{m\over45\big(m(2n+1)+2\big)}\bigg(m^3(4n^2+4n+1)+8m^2(2n+1)\cr
  &+m(4n^4+8n^3-24n^2-28n+4)+16n^3+24n^2+2n-3\bigg).
  }
  $$
Setting $n=1,2,\ldots$ reproduces a subset of our previous expressions. For
example $n=1$ gives (\peq{zag2}). The advantage of this form, and its
companion, (\peq{exeq1}), is the explicit dependence on $n$.

Incidentally, the identity (\peq{cotid}) can be used to obtain the specific
value of the original Dedekind sum, $s(2,q)=(q-1)(q-5)/24q$. More generally
it allows one to obtain explicit forms for the sums,
$\tau(q;1,\ldots,1,2,\ldots,2)$, as combinatorial combinations of
generalised Bernoulli polynomials in $q$ yielding, for example,
  $$\eqalign{
  \tau(q;1,1,1,2,2,2)=-{1\over7560q}&(q-1)(q-5)(2q^4+12q^3-43q^2-318q+995)\cr
  \tau(q;1,1,1,1,2,2)=-{1\over1890q}&(q-1)(q-2)(q-4)(2q^3+7q^2-7q-139)\cr
  \tau(q;1,1,1,1,1,2,2,2,2,2)=&\cr
  -{1\over2993760q}&(q-1)(q-5)(2q^8+12q^7-103q^6-678q^5\cr
  &+2013q^4+15468q^3-18017q^2-185442q+344425)\,.
  }
  $$
Curiously, the symmetrical quantities seem to vanish when $q=5$ on the
spheres S$^3$, S$^{11}$, S$^{19}\,\ldots$ but not on S$^7$,
S$^{15}\,\ldots$.

The identity also permits a sum such as $\tau(q;1,2,\nu,\ldots,\mu)$ to be
replaced by the `simpler' ones, $\tau(q;1,1,\nu,\ldots,\mu)$ and
$\tau(q;\nu,\ldots,\mu)$. More pertinant for this paper, it means that
$\tau(q;1,2,\nu,\nu)$ can be reduced to $\tau(q;1,1,\nu,\nu)$ ($q$ odd),
computed previously, and to $\tau(q;\nu,\nu)=\tau(q;1,1)$, a standard sum,
  $$
  \tau(q;1,2,\nu,\nu)={1\over2}\tau(q,\nu)-{1\over6}(q-1)(q-2)\,.
  $$
The table in Zagier enables one to check this relation, which I would like
to regard as extending the scope of my earlier results.

The consequences of more complicated trigonometric identities are discussed
by Zagier, [\pref{Zagier}]. He shows, for example, that $\tau(q;1,1,2,4)=
\tau(q;1,2,2,2)$ by an identity following from (\peq{cotid}). Actually,
repeated direct application of just (\peq{cotid}) reduces each of these
quantitites ultimately to standard summations, so rendering Zagier's earlier
computation via reciprocity unnecessary.

\newpage

\section{\bf References.}
\begin{putreferences}
  \ref{DandA}{Dowker,J.S. and Apps, J.S. \cqg{}{}{}.}
  \ref{Weil}{Weil,A., {\it Elliptic functions according to Eisenstein
  and Kronecker}, Springer, Berlin, 1976.}
  \ref{Ling}{Ling,C-H. {\it SIAM J.Math.Anal.} {\bf5} (1974) 551.}
  \ref{Ling2}{Ling,C-H. {\it J.Math.Anal.Appl.}(1988).}
 \ref{BMO}{Brevik,I., Milton,K.A. and Odintsov, S.D. {\it Entropy bounds in
 $R\times S^3$ geometries}. hep-th/0202048.}
 \ref{KandL}{Kutasov,D. and Larsen,F. {\it JHEP} 0101 (2001) 1.}
 \ref{KPS}{Klemm,D., Petkou,A.C. and Siopsis {\it Entropy
 bounds, monoticity properties and scaling in CFT's}. hep-th/0101076.}
 \ref{DandC}{Dowker,J.S. and Critchley,R. \prD{15}{1976}{1484}.}
 \ref{AandD}{Al'taie, M.B. and Dowker, J.S. \prD{18}{1978}{3557}.}
 \ref{Dow1}{Dowker,J.S. \prD{37}{1988}{558}.}
 \ref{Dow3}{Dowker,J.S. \prD{28}{1983}{3013}.}
 \ref{DandK}{Dowker,J.S. and Kennedy,G. \jpa{}{1978}{}.}
 \ref{Dow2}{Dowker,J.S. \cqg{1}{1984}{359}.}
 \ref{DandKi}{Dowker,J.S. and Kirsten, K.{\it Comm. in Anal. and Geom.
 }{\bf7}(1999) 641.}
 \ref{DandKe}{Dowker,J.S. and Kennedy,G.\jpa{11}{1978}{895}.}
 \ref{Gibbons}{Gibbons,G.W. \pl{60A}{1977}{385}.}
 \ref{Cardy}{Cardy,J.L. \np{366}{1991}{403}.}
 \ref{ChandD}{Chang,P. and Dowker,J.S. \np{395}{1993}{407}.}
 \ref{DandC2}{Dowker,J.S. and Critchley,R. \prD{13}{1976}{224}.}
 \ref{Camporesi}{Camporesi,R. \prp{196}{1990}{1}.}
 \ref{BandM}{Brown,L.S. and Maclay,G.J. \pr{184}{1969}{1272}.}
 \ref{CandD}{Candelas,P. and Dowker,J.S. \prD{19}{1979}{2902}.}
 \ref{Unwin1}{Unwin,S.D. Thesis. University of Manchester. 1979.}
 \ref{Unwin2}{Unwin,S.D. \jpa{13}{1980}{313}.}
 \ref{DandB}{Dowker,J.S.and Banach,R. \jpa{11}{1978}{2255}.}
 \ref{Obhukov}{Obhukov,Yu.N. \pl{109B}{1982}{195}.}
 \ref{Kennedy}{Kennedy,G. \prD{23}{1981}{2884}.}
 \ref{CandT}{Copeland,E. and Toms,D.J. \np {255}{1985}{201}.}
 \ref{ELV}{Elizalde,E., Lygren, M. and Vassilevich,
 D.V. \jmp {37}{1996}{3105}.}
 \ref{Malurkar}{Malurkar,S.L. {\it J.Ind.Math.Soc} {\bf16} (1925/26) 130.}
 \ref{Glaisher}{Glaisher,J.W.L. {\it Messenger of Math.} {\bf18}
(1889) 1.} \ref{Anderson}{Anderson,A. \prD{37}{1988}{536}.}
 \ref{CandA}{Cappelli,A. and D'Appollonio,\pl{487B}{2000}{87}.}
 \ref{Wot}{Wotzasek,C. \jpa{23}{1990}{1627}.}
 \ref{RandT}{Ravndal,F. and Tollesen,D. \prD{40}{1989}{4191}.}
 \ref{SandT}{Santos,F.C. and Tort,A.C. \pl{482B}{2000}{323}.}
 \ref{FandO}{Fukushima,K. and Ohta,K. {\it Physica} {\bf A299} (2001) 455.}
 \ref{GandP}{Gibbons,G.W. and Perry,M. \prs{358}{1978}{467}.}
 \ref{Dow4}{Dowker,J.S. {\it Zero modes, entropy bounds and partition
functions.} hep-th\break /0203026.}
  \ref{Rad}{Rademacher,H. {\it Topics in analytic number theory,}
Springer-Verlag,  Berlin,1973.}
  \ref{Halphen}{Halphen,G.-H. {\it Trait\'e des Fonctions Elliptiques}, Vol 1,
Gauthier-Villars, Paris, 1886.}
  \ref{CandW}{Cahn,R.S. and Wolf,J.A. {\it Comm.Mat.Helv.} {\bf 51} (1976) 1.}
  \ref{Berndt}{Berndt,B.C. \rmjm{7}{1977}{147}.}
  \ref{Hurwitz}{Hurwitz,A. \ma{18}{1881}{528}.}
  \ref{Hurwitz2}{Hurwitz,A. {\it Mathematische Werke} Vol.I. Basel,
  Birkhauser, 1932.}
  \ref{Berndt2}{Berndt,B.C. \jram{303/304}{1978}{332}.}
  \ref{RandA}{Rao,M.B. and Ayyar,M.V. \jims{15}{1923/24}{150}.}
  \ref{Hardy}{Hardy,G.H. \jlms{3}{1928}{238}.}
  \ref{TandM}{Tannery,J. and Molk,J. {\it Fonctions Elliptiques},
   Gauthier-Villars, Paris, 1893--1902.}
  \ref{schwarz}{Schwarz,H.-A. {\it Formeln und Lehrs\"atzen zum Gebrauche..},
  Springer 1893.(The first edition was 1885.) The French translation by
Henri Pad\'e is {\it Formules et Propositions pour L'Emploi...},
Gauthier-Villars, Paris, 1894}
  \ref{Hancock}{Hancock,H. {\it Theory of elliptic functions}, Vol I.
   Wiley, New York 1910.}
  \ref{watson}{Watson,G.N. \jlms{3}{1928}{216}.}
  \ref{MandO}{Magnus,W. and Oberhettinger,F. {\it Formeln und S\"atze},
  Springer-Verlag, Berlin 1948.}
  \ref{Klein}{Klein,F. {\it Lectures on the Icosohedron}
  (Methuen, London, 1913).}
  \ref{AandL}{Appell,P. and Lacour,E. {\it Fonctions Elliptiques},
  Gauthier-Villars,
  Paris, 1897.}
  \ref{HandC}{Hurwitz,A. and Courant,C. {\it Allgemeine Funktionentheorie},
  Springer,
  Berlin, 1922.}
  \ref{WandW}{Whittaker,E.T. and Watson,G.N. {\it Modern analysis},
  Cambridge 1927.}
  \ref{SandC}{Selberg,A. and Chowla,S. \jram{227}{1967}{86}. }
  \ref{zucker}{Zucker,I.J. {\it Math.Proc.Camb.Phil.Soc} {\bf 82 }(1977) 111.}
  \ref{glasser}{Glasser,M.L. {\it Maths.of Comp.} {\bf 25} (1971) 533.}
  \ref{GandW}{Glasser, M.L. and Wood,V.E. {\it Maths of Comp.} {\bf 25} (1971)
  535.}
  \ref{greenhill}{Greenhill,A,G. {\it The Applications of Elliptic
  Functions}, MacMillan, London, 1892.}
  \ref{Weierstrass}{Weierstrass,K. {\it J.f.Mathematik (Crelle)}
{\bf 52} (1856) 346.}
  \ref{Weierstrass2}{Weierstrass,K. {\it Mathematische Werke} Vol.I,p.1,
  Mayer u. M\"uller, Berlin, 1894.}
  \ref{Fricke}{Fricke,R. {\it Die Elliptische Funktionen und Ihre Anwendungen},
    Teubner, Leipzig. 1915, 1922.}
  \ref{Konig}{K\"onigsberger,L. {\it Vorlesungen \"uber die Theorie der
 Elliptischen Funktionen},  \break Teubner, Leipzig, 1874.}
  \ref{Milne}{Milne,S.C. {\it The Ramanujan Journal} {\bf 6} (2002) 7-149.}
  \ref{Schlomilch}{Schl\"omilch,O. {\it Ber. Verh. K. Sachs. Gesell. Wiss.
  Leipzig}  {\bf 29} (1877) 101-105; {\it Compendium der h\"oheren Analysis},
  Bd.II, 3rd Edn, Vieweg, Brunswick, 1878.}
  \ref{BandB}{Briot,C. and Bouquet,C. {\it Th\`eorie des Fonctions
  Elliptiques}, Gauthier-Villars, Paris, 1875.}
  \ref{Dumont}{Dumont,D. \aim {41}{1981}{1}.}
  \ref{Andre}{Andr\'e,D. {\it Ann.\'Ecole Normale Superior} {\bf 6} (1877) 265;
  {\it J.Math.Pures et Appl.} {\bf 5} (1878) 31.}
  \ref{Raman}{Ramanujan,S. {\it Trans.Camb.Phil.Soc.} {\bf 22} (1916) 159;
 {\it Collected Papers}, Cambridge, 1927}
  \ref{Weber}{Weber,H.M. {\it Lehrbuch der Algebra} Bd.III, Vieweg,
  Brunswick 190  3.}
  \ref{Weber2}{Weber,H.M. {\it Elliptische Funktionen und algebraische Zahlen},
  Vieweg, Brunswick 1891.}
  \ref{ZandR}{Zucker,I.J. and Robertson,M.M.
  {\it Math.Proc.Camb.Phil.Soc} {\bf 95 }(1984) 5.}
  \ref{JandZ1}{Joyce,G.S. and Zucker,I.J.
  {\it Math.Proc.Camb.Phil.Soc} {\bf 109 }(1991) 257.}
  \ref{JandZ2}{Zucker,I.J. and Joyce.G.S.
  {\it Math.Proc.Camb.Phil.Soc} {\bf 131 }(2001) 309.}
  \ref{zucker2}{Zucker,I.J. {\it SIAM J.Math.Anal.} {\bf 10} (1979) 192,}
  \ref{BandZ}{Borwein,J.M. and Zucker,I.J. {\it IMA J.Math.Anal.} {\bf 12}
  (1992) 519.}
  \ref{Cox}{Cox,D.A. {\it Primes of the form $x^2+n\,y^2$}, Wiley, New York,
  1989.}
  \ref{BandCh}{Berndt,B.C. and Chan,H.H. {\it Mathematika} {\bf42} (1995) 278.}
  \ref{EandT}{Elizalde,R. and Tort.hep-th/}
  \ref{KandS}{Kiyek,K. and Schmidt,H. {\it Arch.Math.} {\bf 18} (1967) 438.}
  \ref{Oshima}{Oshima,K. \prD{46}{1992}{4765}.}
  \ref{greenhill2}{Greenhill,A.G. \plms{19} {1888} {301}.}
  \ref{Russell}{Russell,R. \plms{19} {1888} {91}.}
  \ref{BandB}{Borwein,J.M. and Borwein,P.B. {\it Pi and the AGM}, Wiley,
  New York, 1998.}
  \ref{Resnikoff}{Resnikoff,H.L. \tams{124}{1966}{334}.}
  \ref{vandp}{Van der Pol, B. {\it Indag.Math.} {\bf18} (1951) 261,272.}
  \ref{Rankin}{Rankin,R.A. {\it Modular forms} CUP}
  \ref{Rankin2}{Rankin,R.A. {\it Proc. Roy.Soc. Edin.} {\bf76 A} (1976) 107.}
  \ref{Skoruppa}{Skoruppa,N-P. {\it J.of Number Th.} {\bf43} (1993) 68 .}
  \ref{Down}{Dowker.J.S. \np {104}{2002}{153}.}
  \ref{Eichler}{Eichler,M. \mz {67}{1957}{267}.}
  \ref{Zagier}{Zagier,D. \invm{104}{1991}{449}.}
  \ref{Lang}{Lang,S. {\it Modular Forms}, Springer, Berlin, 1976.}
  \ref{Kosh}{Koshliakov,N.S. {\it Mess.of Math.} {\bf 58} (1928) 1.}
  \ref{BandH}{Bodendiek, R. and Halbritter,U. \amsh{38}{1972}{147}.}
  \ref{Smart}{Smart,L.R., \pgma{14}{1973}{1}.}
  \ref{Grosswald}{Grosswald,E. {\it Acta. Arith.} {\bf 21} (1972) 25.}
  \ref{Kata}{Katayama,K. {\it Acta Arith.} {\bf 22} (1973) 149.}
  \ref{Ogg}{Ogg,A. {\it Modular forms and Dirichlet series} (Benjamin,
  New York,
   1969).}
  \ref{Bol}{Bol,G. \amsh{16}{1949}{1}.}
  \ref{Epstein}{Epstein,P. \ma{56}{1903}{615}.}
  \ref{Petersson}{Petersson.}
  \ref{Serre}{Serre,J-P. {\it A Course in Arithmetic}, Springer,
  New York, 1973.}
  \ref{Schoenberg}{Schoenberg,B., {\it Elliptic Modular Functions},
  Springer, Berlin, 1974.}
  \ref{Apostol}{Apostol,T.M. \dmj {17}{1950}{147}.}
  \ref{Ogg2}{Ogg,A. {\it Lecture Notes in Math.} {\bf 320} (1973) 1.}
  \ref{Knopp}{Knopp,M.I. \dmj {45}{1978}{47}.}
  \ref{Knopp2}{Knopp,M.I. \invm {}{1994}{361}.}
  \ref{LandZ}{Lewis,J. and Zagier,D. \aom{153}{2001}{191}.}
  \ref{DandK1}{Dowker,J.S. and Kirsten,K. {\it Elliptic functions and
  temperature inversion symmetry on spheres} hep-th/.}
  \ref{HandK}{Husseini and Knopp.}
  \ref{Kober}{Kober,H. \mz{39}{1934-5}{609}.}
  \ref{HandL}{Hardy,G.H. and Littlewood, \am{41}{1917}{119}.}
  \ref{Watson}{Watson,G.N. \qjm{2}{1931}{300}.}
  \ref{SandC2}{Chowla,S. and Selberg,A. {\it Proc.Nat.Acad.} {\bf 35}
  (1949) 371.}
  \ref{Landau}{Landau, E. {\it Lehre von der Verteilung der Primzahlen},
  (Teubner, Leipzig, 1909).}
  \ref{Berndt4}{Berndt,B.C. \tams {146}{1969}{323}.}
  \ref{Berndt3}{Berndt,B.C. \tams {}{}{}.}
  \ref{Bochner}{Bochner,S. \aom{53}{1951}{332}.}
  \ref{Weil2}{Weil,A.\ma{168}{1967}{}.}
  \ref{CandN}{Chandrasekharan,K. and Narasimhan,R. \aom{74}{1961}{1}.}
  \ref{Rankin3}{Rankin,R.A. {} {} ().}
  \ref{Berndt6}{Berndt,B.C. {\it Trans.Edin.Math.Soc}.}
  \ref{Elizalde}{Elizalde,E. {\it Ten Physical Applications of Spectral
  Zeta Function Theory}, \break (Springer, Berlin, 1995).}
  \ref{Allen}{Allen,B., Folacci,A. and Gibbons,G.W. \pl{189}{1987}{304}.}
  \ref{Krazer}{Krazer}
  \ref{Elizalde3}{Elizalde,E. {\it J.Comp.and Appl. Math.} {\bf 118}
  (2000) 125.}
  \ref{Elizalde2}{Elizalde,E., Odintsov.S.D, Romeo, A. and Bytsenko,
  A.A and
  Zerbini,S.
  {\it Zeta function regularisation}, (World Scientific, Singapore,
  1994).}
  \ref{Eisenstein}{Eisenstein}
  \ref{Hecke}{Hecke,E. \ma{112}{1936}{664}.}
  \ref{Terras}{Terras,A. {\it Harmonic analysis on Symmetric Spaces} (Springer,
  New York, 1985).}
  \ref{BandG}{Bateman,P.T. and Grosswald,E. {\it Acta Arith.} {\bf 9}
  (1964) 365.}
  \ref{Deuring}{Deuring,M. \aom{38}{1937}{585}.}
  \ref{Guinand}{Guinand.}
  \ref{Guinand2}{Guinand.}
  \ref{Minak}{Minakshisundaram.}
  \ref{Mordell}{Mordell,J. \prs{}{}{}.}
  \ref{GandZ}{Glasser,M.L. and Zucker, {}.}
  \ref{Landau2}{Landau,E. \jram{}{1903}{64}.}
  \ref{Kirsten1}{Kirsten,K. \jmp{35}{1994}{459}.}
  \ref{Sommer}{Sommer,J. {\it Vorlesungen \"uber Zahlentheorie}
  (1907,Teubner,Leipzig).
  French edition 1913 .}
  \ref{Reid}{Reid,L.W. {\it Theory of Algebraic Numbers},
  (1910,MacMillan,New York).}
  \ref{Milnor}{Milnor, J. {\it Is the Universe simply--connected?},
  IAS, Princeton, 1978.}
  \ref{Milnor2}{Milnor, J. \ajm{79}{1957}{623}.}
  \ref{Opechowski}{Opechowski,W. {\it Physica} {\bf 7} (1940) 552.}
  \ref{Bethe}{Bethe, H.A. \zfp{3}{1929}{133}.}
  \ref{LandL}{Landau, L.D. and Lishitz, E.M. {\it Quantum
  Mechanics} (Pergamon Press, London, 1958).}
  \ref{GPR}{Gibbons, G.W., Pope, C. and R\"omer, H., \np{157}{1979}{377}.}
  \ref{Jadhav}{Jadhav,S.P. PhD Thesis, University of Manchester 1990.}
  \ref{DandJ}{Dowker,J.S. and Jadhav, S. \prD{39}{1989}{1196}.}
  \ref{CandM}{Coxeter, H.S.M. and Moser, W.O.J. {\it Generators and
  relations of finite groups} Springer. Berlin. 1957.}
  \ref{Coxeter2}{Coxeter, H.S.M. {\it Regular Complex Polytopes},
   (Cambridge University Press,
  Cambridge, 1975).}
  \ref{Coxeter}{Coxeter, H.S.M. {\it Regular Polytopes}.}
  \ref{Stiefel}{Stiefel, E., J.Research NBS {\bf 48} (1952) 424.}
  \ref{BandS}{Brink and Satchler {\it Angular momentum theory}.}
  \ref{Rose}{Rose}
  \ref{Schwinger}{Schwinger,J.}
  \ref{Bromwich}{Bromwich, T.J.I'A. {\it Infinite Series},
  (Macmillan, 1947).}
  \ref{Ray}{Ray,D.B. \aim{4}{1970}{109}.}
  \ref{Ikeda}{Ikeda,A. {\it Kodai Math.J.} {\bf 18} (1995) 57.}
  \ref{Kennedy}{Kennedy,G. \prD{23}{1981}{2884}.}
  \ref{Ellis}{Ellis,G.F.R. {\it General Relativity} {\bf2} (1971) 7.}
  \ref{Dow8}{Dowker,J.S. \cqg{20}{2003}{L105}.}
  \ref{IandY}{Ikeda, A and Yamamoto, Y. \ojm {16}{1979}{447}.}
  \ref{BandI}{Bander,M. and Itzykson,C. \rmp{18}{1966}{2}.}
  \ref{Schulman}{Schulman, L.S. \pr{176}{1968}{1558}.}
  \ref{Bar1}{B\"ar,C. {\it Arch.d.Math.}{\bf 59} (1992) 65.}
  \ref{Bar2}{B\"ar,C. {\it Geom. and Func. Anal.} {\bf 6} (1996) 899.}
  \ref{Vilenkin}{Vilenkin, N.J. {\it Special functions},
  (Am.Math.Soc., Providence, 1968).}
  \ref{Talman}{Talman, J.D. {\it Special functions} (Benjamin,N.Y.,1968).}
  \ref{Miller}{Miller,W. {\it Symmetry groups and their applications}
  (Wiley, N.Y., 1972).}
  \ref{Dow3}{Dowker,J.S. \cmp{162}{1994}{633}.}
  \ref{Cheeger}{Cheeger, J. \jdg {18}{1983}{575}.}
  \ref{Dow6}{Dowker,J.S. \jmp{30}{1989}{770}.}
  \ref{Dow9}{Dowker,J.S. \jmp{42}{2001}{1501}.}
  \ref{Dow7}{Dowker,J.S. \jpa{25}{1992}{2641}.}
  \ref{Warner}{Warner.N.P. \prs{383}{1982}{379}.}
  \ref{Wolf}{Wolf, J.A. {\it Spaces of constant curvature},
  (McGraw--Hill,N.Y., 1967).}
  \ref{Meyer}{Meyer,B. \cjm{6}{1954}{135}.}
  \ref{BandB}{B\'erard,P. and Besson,G. {\it Ann. Inst. Four.} {\bf 30}
  (1980) 237.}
  \ref{PandM}{Polya,G. and Meyer,B. \cras{228}{1948}{28}.}
  \ref{Springer}{Springer, T.A. Lecture Notes in Math. vol 585 (Springer,
  Berlin,1977).}
  \ref{SeandT}{Threlfall, H. and Seifert, W. \ma{104}{1930}{1}.}
  \ref{Hopf}{Hopf,H. \ma{95}{1925}{313}. }
  \ref{Dow}{Dowker,J.S. \jpa{5}{1972}{936}.}
  \ref{LLL}{Lehoucq,R., Lachi\'eze-Rey,M. and Luminet, J.--P. {\it
  Astron.Astrophys.} {\bf 313} (1996) 339.}
  \ref{LaandL}{Lachi\'eze-Rey,M. and Luminet, J.--P.
  \prp{254}{1995}{135}.}
  \ref{Schwarzschild}{Schwarzschild, K., {\it Vierteljahrschrift der
  Ast.Ges.} {\bf 35} (1900) 337.}
  \ref{Starkman}{Starkman,G.D. \cqg{15}{1998}{2529}.}
  \ref{LWUGL}{Lehoucq,R., Weeks,J.R., Uzan,J.P., Gausman, E. and
  Luminet, J.--P. \cqg{19}{2002}{4683}.}
  \ref{Dow10}{Dowker,J.S. \prD{28}{1983}{3013}.}
  \ref{BandD}{Banach, R. and Dowker, J.S. \jpa{12}{1979}{2527}.}
  \ref{Jadhav2}{Jadhav,S. \prD{43}{1991}{2656}.}
  \ref{Gilkey}{Gilkey,P.B. {\it Invariance theory,the heat equation and
  the Atiyah--Singer Index theorem} (CRC Press, Boca Raton, 1994).}
  \ref{BandY}{Berndt,B.C. and Yeap,B.P. {\it Adv. Appl. Math.}
  {\bf29} (2002) 358.}
  \ref{HandR}{Hanson,A.J. and R\"omer,H. \pl{80B}{1978}{58}.}
  \ref{Hill}{Hill,M.J.M. {\it Trans.Camb.Phil.Soc.} {\bf 13} (1883) 36.}
  \ref{Cayley}{Cayley,A. {\it Quart.Math.J.} {\bf 7} (1866) 304.}
  \ref{Seade}{Seade,J.A. {\it Anal.Inst.Mat.Univ.Nac.Aut\'on
  M\'exico} {\bf 21} (1981) 129.}
  \ref{CM}{Cisneros--Molina,J.L. {\it Geom.Dedicata} {\bf84} (2001)
  \ref{Goette1}{Goette,S. \jram {526} {2000} 181.}
  207.}
  \ref{NandO}{Nash,C. and O'Connor,D--J, \jmp {36}{1995}{1462}.}
  \ref{Dows}{Dowker,J.S. \aop{71}{1972}{577}; Dowker,J.S. and Pettengill,D.F.
  \jpa{7}{1974}{1527}; J.S.Dowker in {\it Quantum Gravity}, edited by
  S. C. Christensen (Hilger,Bristol,1984)}
  \ref{Jadhav2}{Jadhav,S.P. \prD{43}{1991}{2656}.}
  \ref{Dow11}{Dowker,J.S. {\it Spherical Universe topology and the Casimir
  effect} hep-th/0404093.}
  \ref{Zagier}{Zagier,D. \ma{202}{1973}{149}}
  \ref{RandG}{Rademacher, H. and Grosswald,E. {\it Dedekind Sums},
  (Carus, MAA, 1972).}
  \ref{Berndt7}{Berndt,B, \aim{23}{1977}{285}.}
  \ref{HKMM}{Harvey,J.A., Kutasov,D., Martinec,E.J. and Moore,G. {\it Localised
  Tachyons and RG Flows}, hep-th/0111154.}
  \ref{Beck}{Beck,M., {\it Dedekind Cotangent Sums}, {\it Acta Arithmetica}
  {\bf 109} (2003) 109-139 ; math.NT/0112077.}
  \ref{McInnes}{McInnes,B. {\it APS instability and the topology of the brane
  world}, hep-th/0401035.}
\end{putreferences}

\bye